\documentclass[twocolumn]{aastex701}

\usepackage{graphicx}

\usepackage{graphicx}
\usepackage{multirow}
\usepackage{comment}
\usepackage{amsmath}

\begin{document}

\title{Investigating H\,$\mathrm{II}$ Regions in the Disk of NGC~7331 with the Circumgalactic H$\alpha$ Spectrograph}

\author[0009-0006-0229-2221]{Nazende ipek Kerkeser}
\affiliation{Steward Observatory, University of Arizona, 933 N. Cherry Avenue, Tucson, AZ 85721, USA}
\email{}

\author[0000-0002-4895-6592]{Nicole Melso}
\affiliation{Steward Observatory, University of Arizona, 933 N. Cherry Avenue, Tucson, AZ 85721, USA}
\affiliation{School of Physics and Astronomy, Rochester Institute of Technology, 84 Lomb Memorial Drive, Rochester, NY
14623, USA}
\email{}

\author{David Schiminovich}
\affiliation{Department of Astronomy, Columbia University, 550 W. 120th Street, MC 5246, New York, NY 10027, USA}
\affiliation{Columbia Astrophysics Laboratory, Columbia University, 550 W. 120th Street, MC 5247, New York, NY 10027, USA}
\email{}

\author[0000-0002-3131-7372]{Erika Hamden}
\affiliation{Steward Observatory, University of Arizona, 933 N. Cherry Avenue, Tucson, AZ 85721, USA}
\email{}

\author[0000-0001-7714-6137]{Meghna Sitaram}
\affiliation{Department of Astronomy, Columbia University, 550 W. 120th Street, MC 5246, New York, NY 10027, USA}
\affiliation{Columbia Astrophysics Laboratory, Columbia University, 550 W. 120th Street, MC 5247, New York, NY 10027, USA}
\email{}

\author[0000-0001-8255-7424]{Ignacio Cevallos-Aleman}
\affiliation{Columbia Astrophysics Laboratory, Columbia University, 550 W. 120th Street, MC 5247, New York, NY 10027, USA}
\affiliation{Department of Physics, Columbia University, 538 W. 120th Street, 704 Pupin Hall, MC 5255, New York, NY 10027, USA}
\email{}

\correspondingauthor{Nicole Melso} 
\email{nxmsps@rit.edu}

\begin{abstract}
We investigate the ionized gas kinematics of H\,\textsc{ii} regions in the disk of NGC~7331 using integral field unit data collected with the Circumgalactic H$\alpha$ Spectrograph (CH$\alpha$S). NGC~7331 is a well-studied nearby galaxy with H\,\textsc{ii} regions resolved by seeing-limited observations, making it ideally suited for this work. The galaxy disk features vigorous star formation, especially in the central ring of starburst activity. We present a catalog of 136 H\,\textsc{ii} regions detected in the SIRTF Nearby Galaxies Survey (SINGS) H$\alpha$ image. Using this refined catalog, we perform aperture photometry on the SINGS narrowband H$\alpha$ images of NGC~7331, extracting the H$\alpha$ luminosity L(H$\alpha$) of these regions. We present corresponding measurements of the average line-of-sight ionized gas velocity dispersion $\sigma$ in these H\,\textsc{ii} regions with CH$\alpha$S.  High-resolution velocity and dispersion maps of the galactic disk are produced from the CH$\alpha$S spectral imaging, selecting spaxels with high signal to noise in order to measure velocity dispersions as low as 12 km s$^{-1}$. Our measurements of the L(H$\alpha$), $\rm \Sigma_{SFR}$ and $\sigma$ in NGC~7331 are consistent with spatially resolved observations of H\,\textsc{ii} regions in large surveys of nearby galaxies. We explore the L(H$\alpha$)$- \sigma$ relationship, identifying turbulent H\,\textsc{ii} regions with nonthermal dispersions likely driven by stellar feedback. The dispersion is correlated with the star formation rate surface density, and using the relation $\rm \sigma \propto \epsilon \Sigma_{SFR}^{\alpha}$, H\,\textsc{ii} regions in NGC~7331 are best fit by $\epsilon = 80$ , $\alpha =0.285$. 
\end{abstract}


\section{Introduction} 
\label{sec:intro}

H\,\textsc{ii} regions 
are vital tracers of recent star formation, and they provide key insight into the interactions between stellar feedback and the interstellar medium (ISM). The ISM across redshifts is supersonically turbulent \citep{Glazebrook2013, Ubler2019, Bacchini2020, Rizzo2024}, driven by a combination of internal and external processes including stellar feedback (winds, supernovae), gravitational and magnetic instabilities, galactic shear, and accretion \citep[and references therein]{Elmegreen2004, Glazebrook2013}. Turbulence in the ISM plays a crucial role in regulating star formation, providing global pressure support that counteracts gravity while also creating perturbations that provoke small-scale collapse \citep{MacLow2004}. 

Turbulent motions in the ISM are observationally probed using the gas velocity dispersion, with many studies finding a positive correlation between gas dispersion ($\sigma$) and the luminosity (L) or star formation rate \citep{Lehnert2009, Lehnert2013, Green2010, Green2014, LeTiran2011, Moiseev2015}. This correlation, known as the L$-$$\sigma$ relation, is expected for many models of ISM turbulence \citep{Lehnert2009, Krumholz2016}. The L$-$$\sigma$ relationship has been studied extensively in high-redshift galaxies and local luminous and ultraluminous infrared galaxies with high star formation rates and large gas dispersions \citep[e.g.,][]{Bellocchi2013, Arribas2014, Green2014, Ubler2019, Perna2022} where gravitational instability, gas transport, and external accretion likely contribute significantly to turbulence \citep{Krumholz2018, Ginzburg2022, Mai2024}. Galaxies with lower star formation rates around a few solar masses per year may straddle the boundary between gravity-driven turbulence and stellar feedback-driven models where the ionized gas dispersions are dominated by the internal motions of the H\,\textsc{ii} regions, highlighting the importance of spatially resolved measurements of the L$-$$\sigma$ relation in this regime at low-redshifts \citep{Krumholz2016}.

NGC~7331 provides an interesting mixture of environments for investigating the relationship between star formation and ISM gas kinematics, aided by its close proximity and extensive observation history. 
Multiwavelength observations have identified distinct morphological features in NGC~7331. The galactic disk is bright in H$\alpha$ emission and hosts a large population of H\,\textsc{ii} regions \citep{Rubin1965, Hodge1983, Marcelin1994, Petit1998}. 
A prominent central ring of dust and gas is seen in CO \citep{Young1982}, H\,\textsc{i} \citep{Bosma1978}, IR \citep{Telesco1982, Regan2004}, and H$\alpha$ \citep{Battaner2003}. This ring exhibits starburst activity that accounts for a significant fraction of the galaxy’s total star formation \citep{Battaner2003, Thilker2007}.
Some kinematic studies of NGC~7331 suggest a counterrotating bulge relative to the disk \citep{prada1996}, and peculiar velocities at the inner boundary of the central ring are consistent with ionized gas inflow \citep{Battaner2003}. A large-scale warp in the H\,\textsc{i} gas distribution \citep{Bosma1978}, an extended distribution of debris/plumes/streams surrounding the galaxy, as well as a a large population of dwarf companions, all suggest a history of mergers and tidal interactions \citep{Ludwig2012, Blauensteiner2017}. Complex velocity structure and morphology, combined with enhanced star formation in NGC~7331, provide an ideal setting for studying how star formation influences and is influenced by gas kinematics.

In this paper, we study the integrated properties of 136 H\,\textsc{ii} regions across the disk of NGC~7331 using integral field spectroscopy (IFS). IFS is a crucial observational technique for efficiently studying large catalogs of H\,\textsc{ii} regions in great detail \citep[e.g.,][]{Sanchez2012, Espinosa-Ponce2020, McLeod2020, Cosens2022, Law2022, Congiu2023, Groves2023, Vaught2024}. At the distance of NGC~7331 (14.5 Mpc; \citealp{Freedman2001}) the instrument spatial resolution of $2\farcs5$ corresponds to approximately 175 pc ($70$ pc arcsec$^{-1}$), allowing for the identification of individual H\,\textsc{ii} regions throughout the disk. The H\,\textsc{ii} regions studied in this work have a radius of $\sim 200$ pc on average. We examine the kinematics of these H\,\textsc{ii} regions, focusing on the relationship between the H$\alpha$ luminosity and ionized gas velocity dispersion ($\sigma$).

This paper is outlined as follows. Section \ref{sec:observations} describes the photometric and spectroscopic datasets and the data reduction process. Section \ref{sec:methods} outlines the selection and characterization of the H\,\textsc{ii} regions in our catalog. Section \ref{sec:results} presents the ionized gas morphology in the disk of NGC~7331 and analyzes the H$\alpha$ L$-$$\sigma$ relation. In Section \ref{sec:discussion}, we interpret our findings in the context of existing
literature and theoretical models. Section \ref{sec:summary} concludes with a summary of key results and directions for future research. 

\section{Observations} 
\label{sec:observations}

This work relies on a combination of photometric and spectroscopic datasets, detailed below. NGC~7331 has a wealth of multiwavelength observations, and many of these datasets have aided our analysis. Here, we map the spatiokinematic structure of ionized gas throughout the galactic disk of NGC~7331 in great detail. 

\subsection{Circumgalactic H$\alpha$ Spectrograph Observations}
Spectroscopic data were collected with the recently commissioned Circumgalactic H$\alpha$ Spectrograph \citep[CH$\alpha$S;][]{Melso2022}. CH$\alpha$S is an advanced IFS optimized for mapping the spatial and kinematic structure of ultrafaint, ionized gas. Accordingly, CH$\alpha$S can easily detect high-surface-brightness H$\alpha$ emission from H\,\textsc{ii} regions in the galactic disk, resolving complex morphology that can be difficult to fully capture with long-slit spectroscopy. En route to ultradeep observations probing the diffuse outskirts of galaxies, CH$\alpha$S will provide detailed spatial and kinematic characterization of the dense ISM. The maps of NGC~7331 presented in this work are an early demonstration of the full observing power of CH$\alpha$S.

CH$\alpha$S operates with a resolving power of $R \sim 10,000$, a field of view (FOV) of $10' \times 10'$, and a spatial resolution of $2\farcs5$. The entrance to the IFS is a microlens array, which segments the telescope focal plane and which leads to the formation of $>$60,000 spectra, each dispersed over a narrow bandpass to avoid overlap. The spatial resolution is set by the pitch of the microlens array, which is slightly larger than the average seeing. 

In Figure \ref{fig:general} we present spectral imaging of NGC~7331 in H$\alpha$ emission. A summary of the CH$\alpha$S observations is given in Table \ref{table:obs}. These data were collected during the Fall of 2023, under very dark, photometric conditions with $<$2$\%$ Moon illumination and $>$125$^{\circ}$ Moon separation. The stack shown in Figure \ref{fig:general} consists of 21, 360s exposures, coaligned and coadded to create a 2 hr integration. The exposure time for each frame is set such that the image does not drift more than half a lenslet at the entrance to the spectrograph. This drift, primarily due to flexure, can shift the image around 10 lenslets (25$''$) in total over multiple hours of exposure. The process for realigning exposures is explained in Section \ref{sec:reduction}.

\begin{figure}[ht!]
\includegraphics[width=0.49\textwidth]{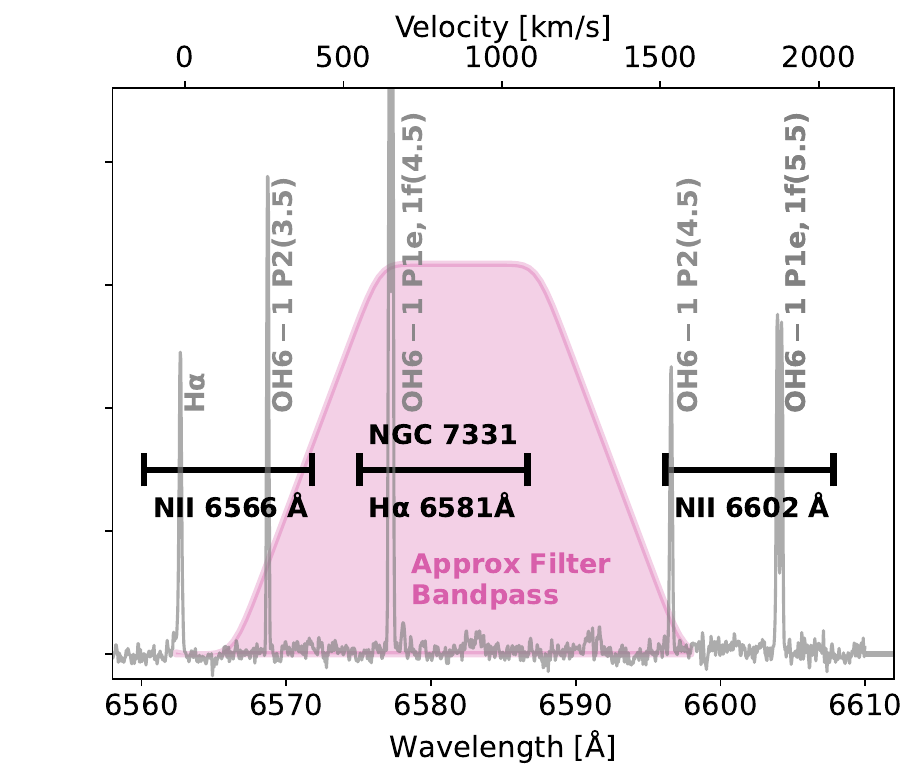}
\caption{We show that H$\alpha$ emission across the velocity profile in the disk of  NGC~7331 is well-covered by the filter, while contamination from [N\,\textsc{ii}] emission is largely rejected. The approximate filter response curve (pink) has a central wavelength of 6582 $\rm \AA$ and a bandpass of 20 $\rm \AA$ FWHM. Black horizontal lines centered on H$\alpha$ and [N\,\textsc{ii}] emission have a width equal to the H\,\textsc{i} 21 cm W$_{20}$ measurement. The sky spectrum (gray) shows the strong Telluric lines; only one bright sky line remains in the filter FWHM \citep{Osterbrock1996}. 
\label{fig:filterband}}
\end{figure}

The narrowband filter combination used for these observations, shown in Figure \ref{fig:filterband}, has a bandpass of 20 $\rm \AA$ FWHM and a central wavelength of 6582 $\rm \AA$. Accordingly, this filter set is best suited for radial velocities ranging from 412 km s$^{-1}$ to 1323 km s$^{-1}$. The systemic velocity of NGC~7331 ($\rm V_{sys} = 816$ km s$^{-1}$, $ \rm \lambda_{sys} = 6581 \ \AA$) is near the center of the filter bandpass, and the galaxy H\,\textsc{i} line full width at $20\%$ of the maximum intensity ($\rm W_{20}$ = 530 km s$^{-1}$) falls within the filter FWHM \citep{Tully1988, Vaucouleurs1991}. This filter combination isolates the H$\alpha$ emission line and rejects contamination from the [N\,\textsc{ii}] doublet (6566 $\rm \AA$, 6602 $\rm \AA$, at the systemic velocity of the galaxy). The bandpass still includes emission from the sky background, notably the bright OH 6-1P1e,1f (4.5) telluric line (unresolved doublet) at 6578 $\rm \AA$ \citep{Osterbrock1996}. See \citealt{Melso2022} for a detailed description of the instrument.

\subsection{Ancillary Data}
This work utilizes multiwavelength observations of NGC~7331, including the narrowband H$\alpha$ imaging from the SIRTF Nearby Galaxies Survey \citep[SINGS;][]{Kennicutt2003, SINGS_DOI}. H\,\textsc{ii} region flux estimates were derived using the corrected SINGS NGC~7331 H$\alpha$ map published in \cite{Leroy2012}. These maps come corrected for [N\,\textsc{ii}] contamination and Galactic extinction, and the integrated flux has been matched to values in the literature. The H\,\textsc{i} 21cm velocity field (moment 1) from The H\,\textsc{i} Nearby Galaxy Survey \citep[THINGS;][]{Walter2008} was used as a reference to create the CH$\alpha$S velocity map. We also compare with the BIMA Survey of Nearby Galaxies (BIMA SONG) CO (1$-$0) map \citep{Helfer2003} in order to identify H\,\textsc{ii} regions located within the gaseous inner ring.

\section{Methods} 
\label{sec:methods}
We detail the methods for extracting kinematic and photometric properties of H\,\textsc{ii} regions from both the CH$\alpha$S and SINGS data. 

\subsection{CH$\alpha$S Data Reduction}
\label{sec:reduction}

\begin{figure*}[ht!]
\centering
\includegraphics[width=\textwidth]{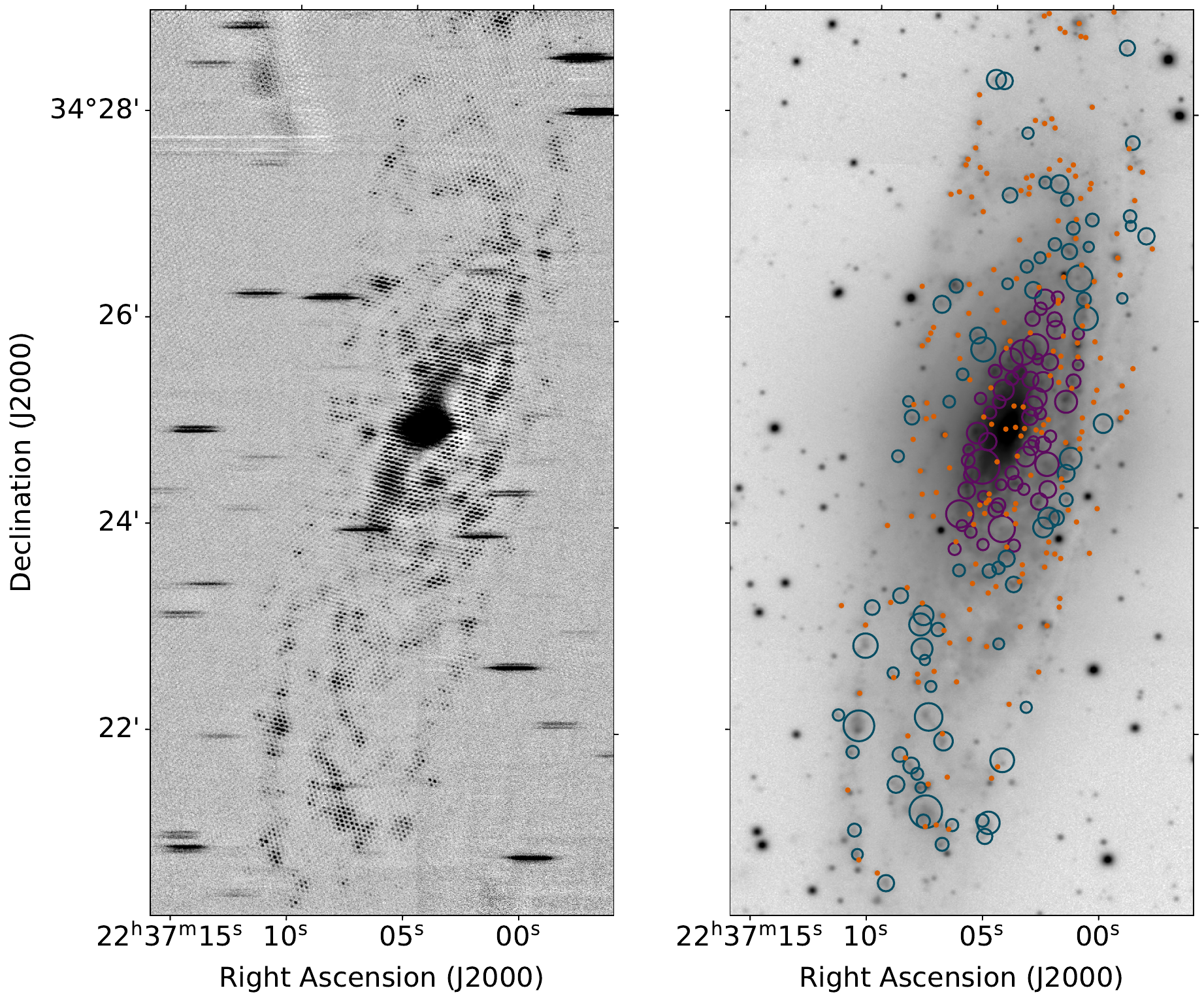}
\caption{Comparison of CH$\alpha$S (left) and SINGS (right) images. The SINGS data is overlaid with our catalog of 136 H\,\textsc{ii} regions represented by circles with diameters corresponding to their measured FWHM. Purple regions are defined as residing in the inner ring selected based on their overlap with the CO ring \citep{Helfer2003}. All other regions shown in green are defined as residing in the outer galactic disk. Regions that have been rejected based on size or
contamination are shown as orange point sources. 
\label{fig:general}}
\end{figure*}

The stacked image shown in Figure \ref{fig:general} is composed of 21 coadded frames, all taken within the same night (see Table \ref{table:obs}). Each frame is bias subtracted before stacking. We do not apply a flat field correction to each individual frame, but we do correct the final stack by an average normalized flat field in order to ensure that relative measurements across the image are consistent. The exposures are aligned and averaged to create the final stack. The drift between exposures is calculated to subpixel precision using the cross correlation in a high signal-to-noise patch of the galaxy. This calculated drift is used in the image registration to shift and align the exposures. The spatial World Coordinate System applied to the CH$\alpha$S data is centered on the wavelength of H$\alpha$ at the systemic velocity of NGC~7331 adjusted for the heliocentric velocity correction. This solution ensures the best average alignment between the CH$\alpha$S spectral imaging in H$\alpha$ emission and narrowband H$\alpha$ imaging from other surveys.

The wavelength/velocity solution is based on the linear dispersion of the instrument. Using a spectral calibration lamp, we create a map of the CH$\alpha$S linear dispersion, which varies across the FOV along the spectral direction. However, because the major axis of NGC~7331 is aligned with the cross-spectral direction, variation in the linear dispersion across the disk is a very small effect on the order of $<$1 km s$^{-1}$ error. We extract and average the linear dispersion values at lenslet positions that cover the disk of NGC~7331, resulting in an average linear dispersion of $\rm 0.37 \ \AA \ pix^{-1}$ or $\rm 16.86 \ km \ s^{-1} \ pix^{-1}$ used in this work.

The stacked CH$\alpha$S data still contains the bright telluric line at 6578 $\rm \AA$. We collect separate sky data at a pointing offset from NGC~7331 and use that sky stack to create the background-subtracted image shown in Figure \ref{fig:general}. To confirm the subtraction accuracy, we measure the residual background counts in blank regions without emission from the galaxy. We compute the summed background counts in a rolling aperture (see Section \ref{sec:apphot}) and examine the distribution; a well-subtracted background results in a histogram with a distribution centered on zero and a standard deviation of less than a few counts. Since the absolute flux measurements are done using the SINGS H$\alpha$ data (see Section \ref{sec:apphot}), we do not apply a flux calibration to the CH$\alpha$S data in this work. The CH$\alpha$S data reduction pipeline is a work in progress \citep[see updates in][]{Cevallos-Aleman2024}.

\subsection{CH$\alpha$S Spectral Extraction and Fitting}
\label{sec:methods-spec}
The CH$\alpha$S FOV consists of $>$60,000 spectra, one for each lenslet in the microlens array. NGC~7331 covers approximately 3000 lenslets, and $\sim 1400$ of the spectra extracted from these lenslets have a a signal-to-noise ratio (S/N) $>3$. We extract each lenslet spectrum using a rectangular aperture centered on the detected H$\alpha$ emission, collapsing each spectrum along the cross-spectral direction to improve the S/N. The mean (velocity centroid) and standard deviation (dispersion) are calculated from the 1D Gaussian fit to the H$\alpha$ emission line in each spectrum. All lines were fit as a single component (one mean velocity peak with a single dispersion value). We calculate the Doppler shift of the H$\alpha$ emission line and the corresponding ionized gas velocity by measuring the offset between the H$\alpha$ emission line peak and the peak emission from the stationary telluric line in the sky background image. The standard deviation of the 1D Gaussian fit to each H$\alpha$ emission line is the observed dispersion ($\sigma_{obs})$. We subtract the instrument dispersion ($\sigma_{inst}$) from the observed dispersion ($\sigma_{obs}$) in quadrature. We similarly correct for thermal broadening at a gas temperature of $10^{4}$ K ($\sigma_{b} = 9.1 \rm \ km \ s^{-1}$) \citep{Rozas2000}. The line-of-sight ionized gas dispersion calculation used in this work follows Equation \ref{eq:disp}:

\begin{equation}
\label{eq:disp}
\sigma = \sqrt{(\sigma_{obs})^{2} - (\sigma_{inst})^{2} - (\sigma_{b})^{2}}
\end{equation}

\begin{figure}[ht!]
\includegraphics[width=0.48\textwidth]{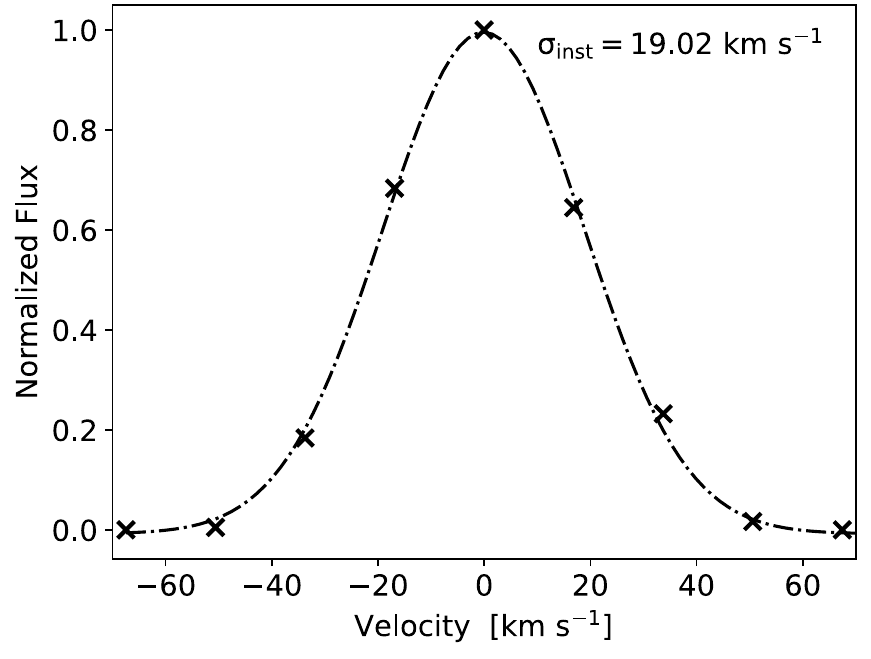}
\caption{ CH$\alpha$S instrument line spread function derived from the stacked sky background line in more than $3000$ spectra. The instrument dispersion measured from the Gaussian fit shown here is convolved with the (much narrower) intrinsic width of the OH 6-1P1e,1f (4.5) telluric line.
\label{fig:sigmainst}}
\end{figure}

We note that in theory we should also correct for the natural line width of H$\alpha$ emission ($\sigma_{N} = 3 \rm \ km \ s^{-1}$); however, the instrument dispersion is measured by stacking the sky spectra extracted from each lenslet and fitting the OH 6-1P1e,1f (4.5) telluric line with a 1D Gaussian. Accordingly, the measured instrument dispersion is convolved with the intrinsic width of the OH 6-1P1e,1f (4.5) telluric line. We assume that the intrinsic width of this telluric line is on the order of the natural line width of H$\alpha$ emission and is already accounted for in our correction for the instrument line spread function. The instrument line spread function and corresponding instrument dispersion derived from a 1D Gaussian fit to the stacked sky background line are shown in Figure \ref{fig:sigmainst}. While fitting calibration lamp data would be higher signal to noise, a single lamp exposure underestimates the instrument dispersion as it does not include the jitter introduced in the observation and coaddition.

To ensure small velocity dispersions below the instrument resolution are measured reliably, we select a signal-to-noise cut on observed flux values (S/N$_{obs}$) such that intrinsic dispersion measurements ($\sigma$) are made with S/N = $\sigma/\delta \sigma = 3$. The observed signal-to-noise (S/N$_{obs}$) required to reach an intrinsic dispersion measurement with an S/N of 3 is given by Equation \ref{eq:snr} (derived in \citealt{Zhou2017}). 

\begin{equation} 
\rm (S/N)_{obs} = (S/N)\left(\sigma_{inst}^{2}/\sigma^{2} + 1\right)
\label{eq:snr}
\end{equation}

Measuring gas dispersions dominated by thermal broadening and natural line width ($\rm \sigma = \sim 12 \ km \ s^{-1}$) with an S/N of 3 requires an observational cut of S/N$_{obs}= 3(\frac{19.02^{2}}{12^{2}} + 1) \approx 10$. A conservative noise estimate is determined by selecting regions in the outskirts of the background-subtracted image and creating a histogram of summed counts in a rolling rectangular aperture with dimensions (w,h) = (5,4) pixels. This aperture size corresponds to emission with a velocity width of $64 \rm \ km \ s^{-1}$, on par with the broadest spectra seen in the disk NGC~7331. The signal is extracted in identical apertures. While the maps presented in Figure \ref{fig:kinematics1} and Figure \ref{fig:kinematics2} use an observed signal-to-noise cut of S/N$> 3$, the dispersion measurements in Figure \ref{fig:Lsigma} and Table \ref{table:HII} use an observed signal to noise cut of S/N$>10$ in order to achieve a $3 \sigma$ measurement of dispersions below the instrument resolution (down to 12 $\rm km \ s^{-1}$).

\begin{deluxetable}{lll}
\tablecaption{NGC~7331 CH$\alpha$S observational summary\label{table:obs}}
\tablehead{
\colhead{Parameter} & \colhead{Value} & \colhead{Comment}
}
\startdata
\textbf{Observation Summary} &  &  \\
R.A. ($\alpha$) & 22:37:04 & FOV center\\
Decl. ($\delta$) & +34:24:56 & FOV center \\
Distance & 14.5 Mpc & \\
Systemic (V$_{\mathrm{Hel}}$) & 816 $\pm$ 1 km s$^{-1}$ & \\
Exposure time & 360 s & Per frame\\
Total integration  & 2.1 hr  & 21 frames \\ 
Moon illumination & $< 2\%$ &  \\
Moon separation & $> 125^{\circ}$ &  \\
Dates &  $19$ Oct 2023 & \\
Heliocentric correction & $29.54$ km s$^{-1}$ & \\
\hline
\textbf{Instrument Settings} &  &  \\
FOV & $10' \times 10'$ & \\
Spatial resolution & $2\farcs8$ & \\
Spectral resolution $(\Delta \lambda)$ & $0.67\,\mathrm{\AA}$ & \\
Central wavelength $(\lambda)$ &  $6582\, \mathrm{\AA}$ &\\ 
Bandpass (FWHM) & $20\, \mathrm{\AA}$ & \\
\enddata
\end{deluxetable}

\subsection{HII Region Selection}
The full catalog of H\,\textsc{ii} regions used in this work is published as a supplementary file, with example entries shown in Table \ref{table:HII}. Performing the H\,\textsc{ii} region detection directly on the CH$\alpha$S spectral image would require smoothing the data, likely overestimating the FWHM fit to these regions and therefore the aperture used for photometry. Instead, the H\,\textsc{ii} regions selected were detected directly in the SINGS image using SExtractor \citep{Bertin1996}. We retain H\,\textsc{ii} regions detected with an S/N$~\geq~$5 over a minimum area of three contiguous pixels. These detection limits return a catalog of 136 H\,\textsc{ii} regions, the majority of which correspond with H\,\textsc{ii} regions identified in the \citealt{Petit1998} [P98] catalog, the \citealt{Marcelin1994} catalog, and the \citealt{Hodge1983} [HK83] catalog of H\,\textsc{ii} regions in NGC~7331. We apply a spatial resolution cut to our catalog, keeping only regions that have an FWHM of $>4''$. The H\,\textsc{ii} regions that remain after this cut are also spatially resolved in the CH$\alpha$S data, covering about two lenslets. These regions do not require aperture corrections, as they are resolved by the seeing-limited PSF ($\sim 1''$$-$$2''$) of the SINGS H$\alpha$ images \citep{Murphy2018}. We visually inspect all apertures and remove regions that are contaminated by their proximity to known stars or are impacted by artifacts such as saturated or oversubtracted pixels. The final catalog is overlaid on Figure \ref{fig:general}. Regions that have been rejected based on size or contamination are shown as orange point sources.

\subsection{SINGS Aperture Photometry}
\label{sec:apphot}
We perform aperture photometry to extract flux values from the SINGS data in the apertures shown in Figure \ref{fig:general}. We chose an aperture diameter corresponding to the FWHM calculated by SExtractor, assuming a Gaussian profile for the region \citep{Bertin1996}. The calibration uncertainty is taken to be $\approx 10\%$ of the flux (DR5 Data Delivery Document). The calibration uncertainty dominates over the background error estimated near the edge of the image, so we assume a $10\%$ flux error on the values quoted in Table \ref{table:HII}. 

\section{Results} 
\label{sec:results}
We report on the ionized gas morphology, luminosity, and kinematics in the disk of NGC~7331. We use these measurements to explore the luminosity$-$velocity dispersion relation for our sample of H\,\textsc{ii} regions described above. 

\subsection{Ionized Gas Morphology}
\label{sec:morphology}
As shown in Figure \ref{fig:general} the ionized gas morphology seen in the CH$\alpha$S spectral imaging is consistent with the structure seen in the SINGS narrowband imaging. Stars in the CH$\alpha$S image are dispersed and appear as bright continuum spectra shortened by the narrow bandpass filter. The LINER nucleus \citep{Cowan1994} is surrounded by faint H$\alpha$ emission, which abruptly increases in brightness due to starburst activity in the inner ring \citep{Battaner2003}. This bright H$\alpha$ emission is coincident with H\,\textsc{i} \citep{Bosma1978} and CO \citep{Young1982} observations of the ring. The outer disk of the galaxy is also bright in discrete H$\alpha$ emission. The H\,\textsc{ii} regions in our catalog are spread throughout the galactic disk. They range in size (FWHM diameter) from approximately 300 pc to 900 pc. We note that this is not an exhaustive catalog. We leave for future work an algorithmic detection of H\,\textsc{ii} regions \citep[e.g.,][]{Sanchez2012, Espinosa-Ponce2020, Congiu2023}, which could be applied to a larger sample of galaxies observed during the CH$\alpha$S commissioning and early science campaigns.

\begin{figure*}[ht!]
\centering
\includegraphics[width=\textwidth]{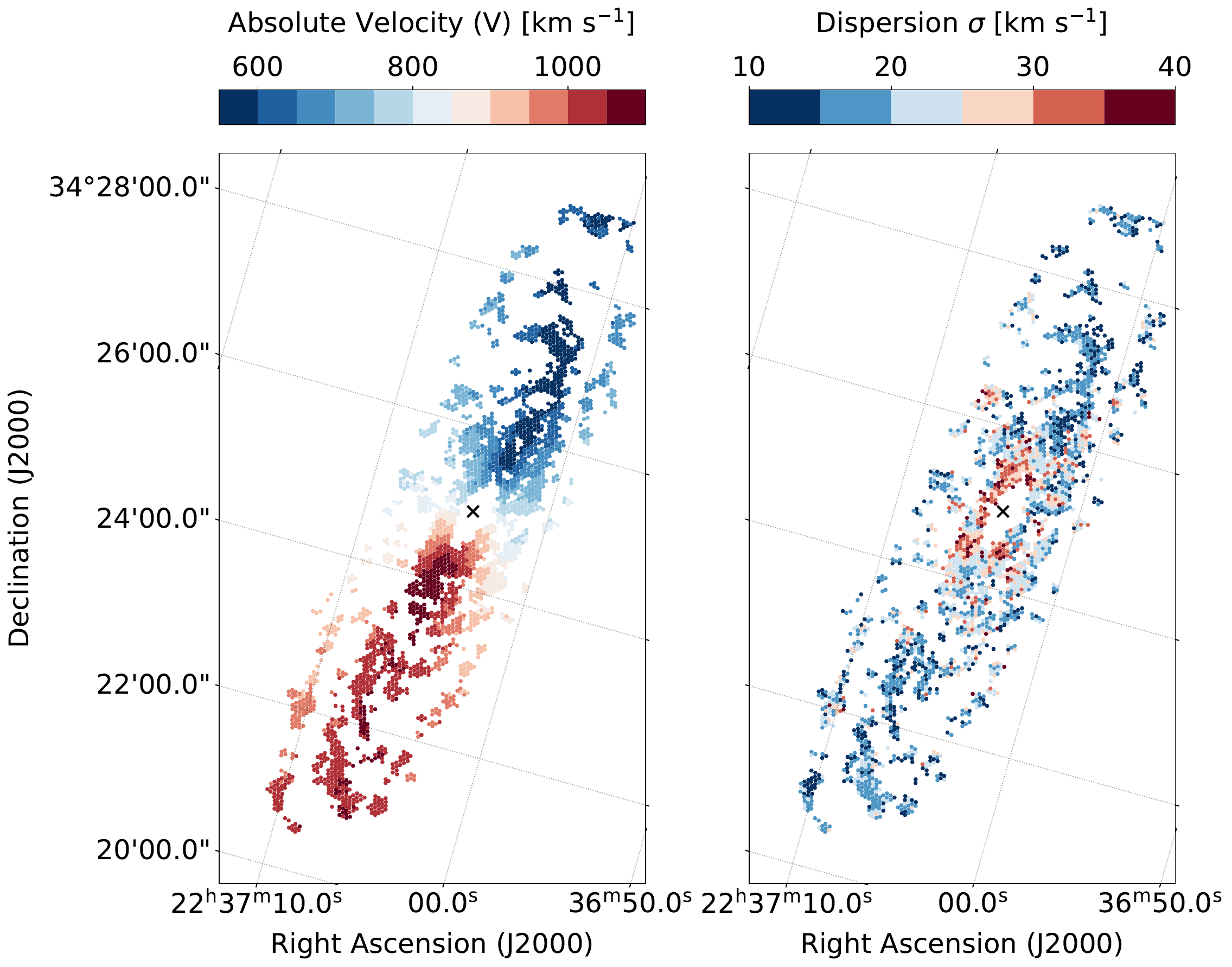}
\caption{Kinematic measurements derived from the CH$\alpha$S IFS. In all panels, the coordinate system has been rotated ($15^{\circ}$) in order to align the
hexagonally packed spectra along the Cartesian y-axis. The transformed coordinates can be compared directly with Figure \ref{fig:general}. We apply an S/N cut on the observed flux, only displaying detections in lenslets with S/N $\geq 3$. The X marker denotes the galactic center coordinates. The individual panel descriptions are as follows: (left) absolute velocity measured in each CH$\alpha$S lenslet, (right) line-of-sight ionized gas velocity dispersion measured in each CH$\alpha$S lenslet. 
\label{fig:kinematics1}}
\end{figure*}

\begin{figure*}[ht!]
\centering
\includegraphics[width=\textwidth]{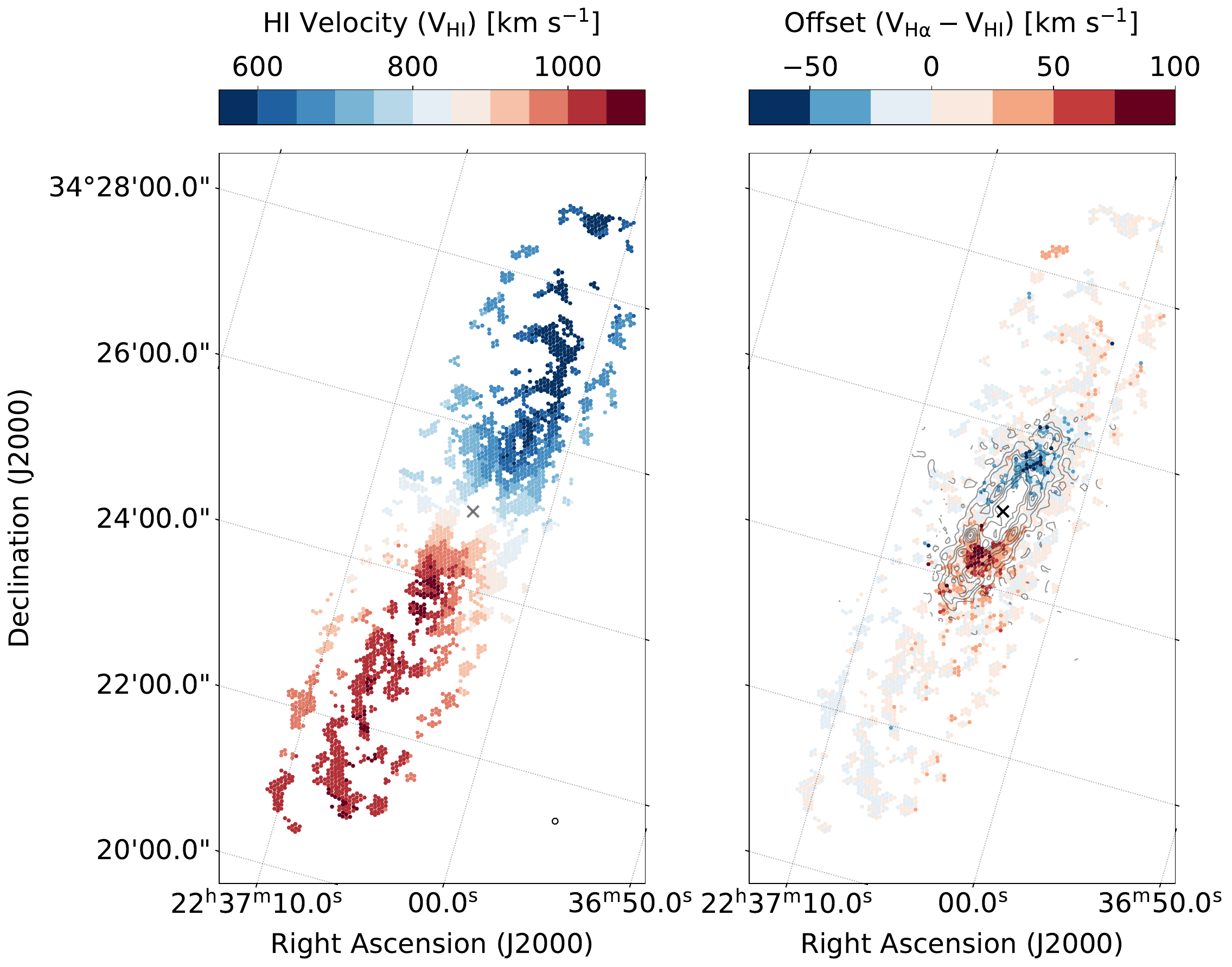}
\caption{Kinematic measurements in NGC~7331. In all panels, the coordinate system has been rotated ($15^{\circ}$) in order to align the
hexagonally packed spectra along the Cartesian y-axis. The transformed coordinates can be compared directly with Figure \ref{fig:general}. We apply an S/N cut, only displaying detections in lenslets with S/N $\geq 3$. The X marker denotes the galactic center coordinates. The individual panel descriptions are as follows: \textit{(left)} H\,\textsc{i} 21cm moment 1 velocity field from THINGS \citep{Walter2008}. This data has been extracted in regions corresponding to each CH$\alpha$S lenslet. The beam size (robust weighting; $\rm B_{maj} = 4.94''$, $B_{min} = 4.60''$ \citealt{Walter2008}) is shown in the lower right corner. \textit{(right)} Residual offset between the H$\alpha$ velocity map and the H\,\textsc{i} velocity map. The gray contours correspond to the inner gas ring seen in the BIMA SONG CO (1$-$0) intensity map. 
\label{fig:kinematics2}}
\end{figure*}

\subsection{Ionized Gas Kinematics}
\label{sec:kinematics}
Maps of the ionized gas velocity and dispersion are made following the spectral fitting procedure described in Section \ref{sec:methods-spec}. The ionized gas velocity map shown in the left panel of Figure \ref{fig:kinematics1} recovers the expected clockwise rotation of the galactic disk seen in H\,\textsc{i} \citep{Walter2008, Schmidt2016}. This velocity map is well matched to prior measurements of the ionized gas velocity traced by Doppler-shifted H$\alpha$ emission shown in \cite{Marcelin1994} and \cite{Daigle2006}. We extend the velocity map to include the high-velocity (red-shifted) edge of the galactic disk that is out of the bandpass in previous datasets. 

We compare the ionized gas velocity with the neutral gas velocity measured in H\,\textsc{i} 21 cm emission. The H\,\textsc{i} velocity field extracted in each lenslet is shown in the left panel of Figure \ref{fig:kinematics2}. In the right panel of Figure \ref{fig:kinematics2} we calculate the residual offset between the ionized gas velocity measured from H$\alpha$ emission and the neutral gas velocity measured from H\,\textsc{i} 21cm emission. In order to interpolate over nan values in the larger beam size of the H\,\textsc{i} velocity moment 1 map (robust weighting; $\rm B_{maj} = 4.94''$, $B_{min} = 4.60''$ \citealt{Walter2008}), we apply small-scale median filtering and Gaussian convolution. As a result, the H\,\textsc{i} velocity map is slightly smoothed. A few spurious values remain in the final H\,\textsc{i} velocity map that propagate to the residual, but these are easily disregarded when examining the two maps by eye. Small residuals on the order of $\pm 25 \rm \ km \ s^{-1}$ (one pixel in the CH$\alpha$S spectra) are within our absolute velocity error bars. Larger residuals, especially those that are spatially coherent over many lenslets, are likely the result of resolution differences (beam smearing) or variation in the neutral and excited gas distributions; however, follow-up analysis is needed to determine if these features may be due to nonrotational, peculiar gas motions.

The line-of-sight ionized gas velocity dispersion ($\sigma$) is shown in the right panel of Figure \ref{fig:kinematics1}. We correct the measured dispersion for the instrument dispersion and thermal broadening (Section \ref{sec:methods-spec}). The remaining quantity is the non-thermal line-of-sight ionized gas velocity dispersion referred to as velocity dispersion throughout the rest of the text. Since all spectra were fit as a single emission line component, this analysis does not account for nonsymmetric features in the line profiles or for multiple peaks from additional resolved velocity components. The average velocity dispersion across all measured spaxels (with an S/N cut of 10) is $20.6 \pm 4.7 \rm \ km \ s^{-1}$, and the (slightly lower) median dispersion across all spaxels is $19.7 \rm \ km \ s^{-1}$. In the outskirts of the galactic disk, measured velocity dispersions are at or near the thermal dispersion limit. We discuss sources of confusion and contamination in these kinematic maps in Section \ref{sec:caveats}.

\subsection{Luminosity-Dispersion Relation}
\label{sec:l-sigma}

\begin{figure}[ht!]
\includegraphics[width=0.465\textwidth]{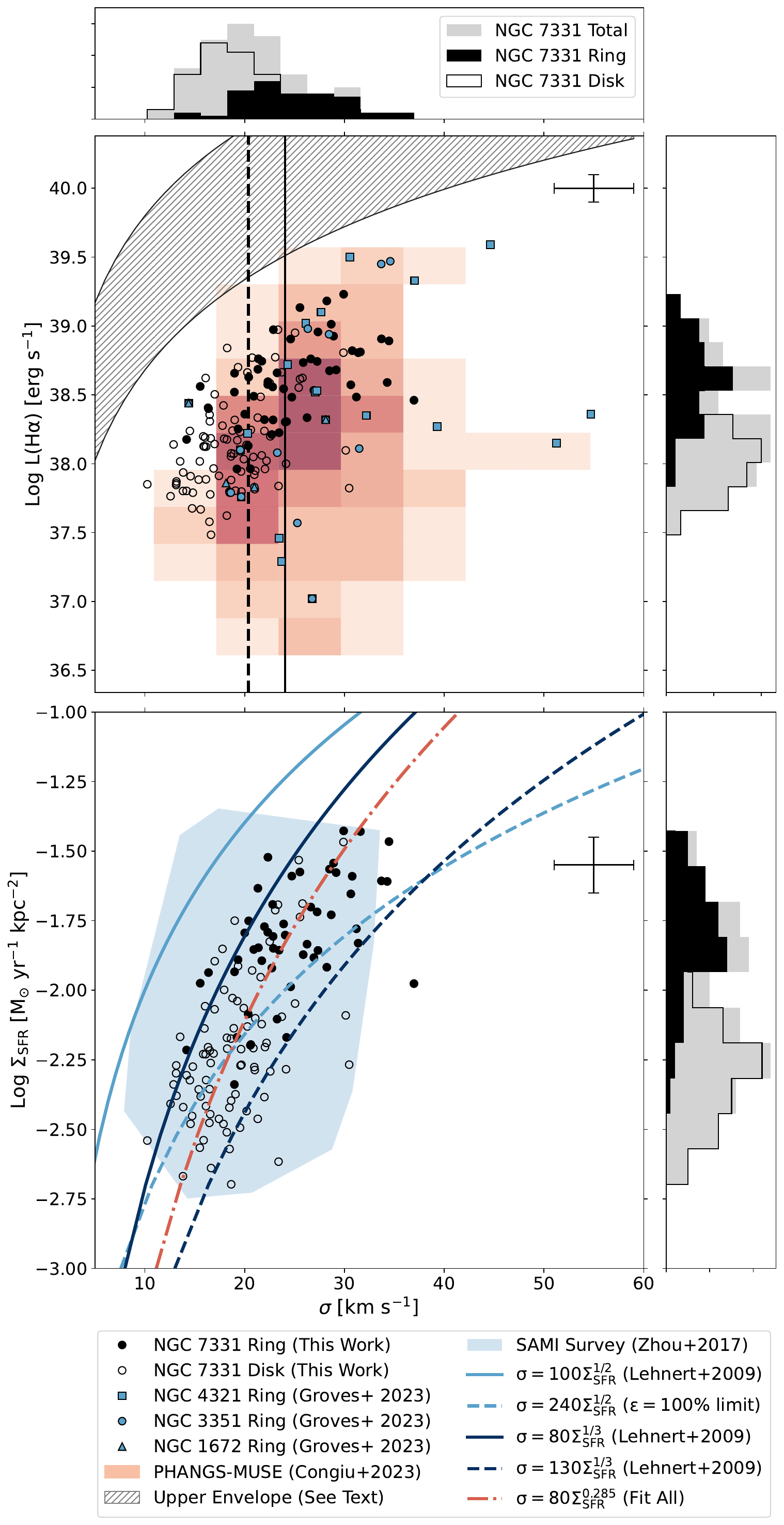}
\caption{Average ionized gas velocity dispersion in extragalactic H\,\textsc{ii} regions as a function of their H$\alpha$ luminosity \textit{(top)} and SFR surface density \textit{(bottom)}. The black points represent the H\,\textsc{ii} regions in NGC~7331 from this work, and they are divided into two populations: those in the nuclear star-forming ring (filled) and those distributed throughout the rest of the galactic disk (open). Corresponding histograms for the disk, ring, and total populations are shown along the sides. We compare our catalog with H\,\textsc{ii} regions in the PHANGS-MUSE catalog (red 2D histogram) \citep{Congiu2023} and the SAMI Galaxy Survey (light blue shading) \citep{Zhou2017}. Models shown are from \citep{Lehnert2009}. See text for more details.
\label{fig:Lsigma}}
\end{figure}

In Figure \ref{fig:Lsigma} we show the observed L$-$$\sigma$ relation for our selected H\,\textsc{ii} regions in NGC~7331, plotting the H$\alpha$ luminosity of the H\,\textsc{ii} regions as a function of their average ionized gas velocity dispersion. We only include H\,\textsc{ii} regions with nonthermal motions, meaning those above a threshold of $\rm \sigma > \sqrt{\sigma_{N}^{2} + \sigma_{b}^{2}}$ or $\rm \sigma > 12 \ km \ s^{-1}$. We calculate both the average velocity dispersion and the average velocity dispersion weighted by the H$\alpha$ intensity. However, since the H$\alpha$ intensity within each region does not vary drastically, this weighting does not change the result significantly, and the average velocity dispersion shown in Figure \ref{fig:Lsigma} is unweighted. 

The H$\alpha$ luminosity values for H\,\textsc{ii} regions measured in NGC~7331 range from $\rm 37.73 <Log(L_{H\alpha}) < 39.16$. These values are uncorrected for intrinsic extinction or diffuse ionized gas (DIG) as discussed in Section \ref{sec:caveats}. The median luminosity of $1\times 10^{38}$ erg  s$^{-1}$ corresponds to $\rm Q_{0} = 7.7 \times 10^{49}$ ionizing photons, equivalent to a minimum of two O5 stars, six O7 stars, or 21 O9 stars (Table 2.3 in \citealt{Osterbrock1974}). Accordingly, the median stellar mass of these regions ranges from 100 to 1000 M$_{\odot}$ (assuming an O star mass of 50 M$_{\odot}$). H$\alpha$ luminosity measurements are converted to a SFR using the prescription in \citealt{Calzetti2013}. The SFR surface density is derived using the area of each H\,\textsc{ii} region calculated from the half-width half-max radius. 

The H\,\textsc{ii} region properties measured in NGC~7331 are consistent with spatially resolved observations of H\,\textsc{ii} regions in large surveys of nearby galaxy disks. In the top panel of Figure \ref{fig:Lsigma} we compare with H\,\textsc{ii} regions in the PHANGS–MUSE nebular catalog \citep{Congiu2023}. Similar to NGC~7331, a few galaxies in the PHANGS-MUSE survey also have nuclear star forming rings \citep{Groves2023}. The blue markers overplotted are H\,\textsc{ii} regions in the nuclear star-forming ring of NGC 4321, NGC 3351, and NGC 1672. In order to ensure a direct comparison, we subtract the natural line width and thermal broadening (in quadrature) from the MUSE H$\alpha$ dispersion measurements and apply a radius cut of $2''$ to match the resolution of our data. After this cut, 964 H\,\textsc{ii} regions remain in the PHANGS-MUSE catalog. The L$-$$\sigma$ distribution of H\,\textsc{ii} regions in the PHANGS-MUSE catalog is shown as a red 2D histogram. In the bottom panel, we compare the SFR surface density of each region with gas dispersions from the SAMI Galaxy Survey shown in light blue \citep{Zhou2017}. We similarly subtract thermal broadening and natural line width in quadrature from these dispersion measurements to make a direct comparison. The SAMI observations already have a reasonably well-matched spatial resolution of $2\farcs5$, so no resolution cut is applied. 

We separate H\,\textsc{ii} regions in NGC~7331 into two populations: those in the nuclear star-forming ring and those distributed throughout the rest of the galactic disk. The boundary of the inner ring used in this work is determined from the BIMA CO intensity map and is defined as an ellipse centered on ($\alpha$,$\delta$) = (22$^{\rm h}$37$^{\rm m}$3.9$^{\rm s}$, +34$^{\rm d}$24$^{\rm m}$55.4$^{\rm s}$) with a major axis $2a = 150''$ and a minor axis $2b =60''$ at an angle of -16$^{\circ}$(clockwise). Bright regions are distributed throughout the disk, but almost all H\,\textsc{ii} regions in the inner ring occupy the high end of the luminosity distribution. There is a trend of high luminosity H\,\textsc{ii} regions in the inner ring having elevated velocity dispersions when compared with H\,\textsc{ii} regions in the rest of the galactic disk. We discuss processes that may be driving these high-velocity dispersions in the inner ring of NGC~7331 in Section \ref{sec:discussion}. 

\section{Discussion} 
\label{sec:discussion}
Our data are consistent with measurements from large surveys of H\,\textsc{ii} regions, including the PHANGS-MUSE Survey \citep{Congiu2023} and the SAMI Galaxy Survey \citep{Zhou2017}. Here, we discuss sources of turbulence driving the velocity dispersions measured in our sample of H\,\textsc{ii} regions alongside caveats to these measurements and potential avenues for future work. 

\subsection{Turbulence Drivers}
\label{subsec:turb}
Past studies of giant H\,\textsc{ii} regions have looked for an envelope to the L$-$$\sigma$ relation or the lowest nonthermal dispersion at a given luminosity \citep{Terlevich1981, Arsenault1990, Relano2005, Zaragoza-Cardiel2017}. We show a range of models (hatched shading) for this envelope in the top panel of Figure \ref{fig:Lsigma}. Regions along this envelope are density bounded, their velocity dispersions are primarily driven by virial motions, and their masses can be approximated from the virial theorem \citep{Rozas1998, Beckman2000, Relano2005, Rozas2006, Blasco-Herrera2010}. A small subset of the most luminous, early-stage H\,\textsc{ii} regions near virial equilibrium may lie on or near the L$-$$\sigma$ envelope. These regions have gravity-driven turbulence. Regions far from this envelope have kinematics dominated by contributions from additional processes. We find that, while some of the brightest H\,\textsc{ii} regions in NGC~7331 may fall on/near the L$-$$\sigma$ envelope after correcting for intrinsic extinction (see Section \ref{sec:caveats}), the majority of H\,\textsc{ii} regions in the disk and ring appear to be driven by alternate processes. This is not surprising, as turbulence driven by gravitational collapse requires massive giant molecular clouds with Jeans masses on the order of $10^{8} \rm M_{\odot}$ and produces relatively low velocity dispersions with an upper limit on the order of $15 \rm \ km \ s^{-1}$ over our range in $\rm \Sigma_{SFR}$ \citep{Lehnert2009, Zhou2017}. Accordingly, we explore other sources of energy injection.

If intense star formation is injecting energy into the ISM, turbulence is likely a combination of larger-scale bulk motions (outflows/shocks that accelerate the surrounding material) and smaller-scale random motions (energy transferred to dense regions and eventually dissipated) \citep{Lehnert2009}. If turbulence is driven by star formation feedback, then the energy injected per unit area should be correlated with the SFR surface density $\sigma \propto \epsilon (\Sigma_{\rm SFR})^{\alpha}$ \citep{Lehnert2009, Green2010, LeTiran2011, Swinbank2012, Lehnert2013, Green2014, Moiseev2015}. In the bottom panel of Figure \ref{fig:Lsigma} we compare the observations with energy injection models formulated as a simple scaling relationship $\sigma \propto (\epsilon \dot{E})^{\alpha}$ \citep{Dib2006, Lehnert2009}. Here, $\epsilon$ is the coupling efficiency of the injected energy transferred to the ISM. For dispersions driven by bulk motions such as supernova explosions, this proportion goes as $\rm \sigma \propto \epsilon \Sigma_{SFR}^{1/2}$. Here the coupling factor for energy injected into the ISM ranges from $\epsilon = 100-240$ for efficiencies of $25\% - 100\%$ \citep{Dib2006}. If instead energy from star formation is dissipated as turbulence dominated by random motions, a steeper model $\rm \sigma \propto \epsilon \Sigma_{SFR}^{1/3}$ has been proposed to fit the dispersion. Here $\epsilon = 80-130$ on 1 kpc injection scales for coupling efficiencies of $25\% - 100\%$ \citep{Lehnert2009}. The steeper model ($\alpha = 1/3$) provides a reasonably good fit to most regions, though some of the H\,\textsc{ii} regions measured require an unphysically large coupling efficiency of $100\%$ in either model. 

Constraining the coupling efficiency to $80 < \epsilon < 240$, we find a best-fit relationship to all H\,\textsc{ii} regions in NGC~7331 of $\rm \sigma \propto 80 \Sigma_{SFR}^{0.285}$. When we separate the disk and ring populations, the fit for H\,\textsc{ii} regions in the ring reveals a slightly shallower relationship than the fit for H\,\textsc{ii} regions in the disk. The shallower relationship in the ring could suggest bulk motions contribute to the dispersion; however, due to scatter in these measurements, this measured variation is likely not significant. Many studies have found a similar power-law relation with $1/3 < \alpha < 1/2$ \citep{Lehnert2009, Zhou2017, Patricio2018, Yu2019, Cui2024}. Power-law values within this range are notably lower than the $\sigma-\Sigma$ relationship expected from the Kennicutt$-$Schmidt scaling law, which for a marginally stable disk (Q $\sim$ 1) predicts $\sigma \propto \Sigma^{0.7}_{\rm SFR}$ \citep{Toomre1964, Kennicutt1998, Krumholz2010, Krumholz2012, Swinbank2012}. 

Increased velocity dispersions in the center of galaxies including our own Milky Way is sometimes attributed to shear/differential rotation in the disk; however, this process increases the dispersion while suppressing star formation, a scenario that is unlikely to explain the increased dispersion seen in the inner ring of NGC~7331 where star formation is enhanced \citep{Krumholz2017, Federrath2016, Kruijssen2017, Federrath2017}. We note that shear may still be contributing in regions with high dispersion and low $\rm \Sigma_{SFR}$. In our catalog, only about $10\%$ of H\,\textsc{ii} regions are both above the median in dispersion and below the median in $\Sigma_{\rm SFR}$.  The radial transport of gas through the disk can also drive turbulent gas motions \citep{Krumholz2018}. However, without a wider range in star formation rate, turbulence driven by gas transport is not easily differentiated from turbulence driven by a combination of feedback and transport \citep{Krumholz2018, Varidel2020}. Further modeling of the disk rotation is needed to compare with the observed velocity map and assess the degree of ionized gas transport in the disk of NGC~7331 \citep{Mai2024}. 

\subsection{Residual Kinematic Offsets}
\label{sec:offsets}
As seen in Figure \ref{fig:kinematics2}, we explore the residual offset between the ionized H$\alpha$ gas velocity and the neutral H\,\textsc{i} gas velocity. Similar to \citealt{Phookun1993} and \citealt{Mitchell2015}, these velocity residuals are patchy and not well correlated with the spiral arms. Prominent differences in the residual velocity map include the redshifted patch in the northern outer disk and the biconical region of high-velocity residual in the galaxy center (CO ring contours overlaid on Figure \ref{fig:kinematics2}).

The redshifted patch seen in the northern galactic disk may be an ionized cloud of inflowing gas seen in projection against the H\,\textsc{i} rotation curve of the galaxy. A few H\,\textsc{ii} regions in NGC 1365 have ionized gas velocities that differ from the H\,\textsc{i} gas velocity by 60-80 $\rm km \ s^{-1}$, attributed to infalling gas clouds beyond the plane of the disk viewed in projection \citep{ZanmarSanchez2008}. The H\,\textsc{i} map of NGC~7331 shows an extended northern lobe of neutral gas where the H\,\textsc{i} gas velocity suggests localized gas inflow may be occurring  \citep{Schmidt2016}. If present, ionized gas inflow in combination with the H\,\textsc{i} inflow rate of 1M$_\odot$ yr$^{-1}$ \citep{Schmidt2016} may be able to sustain the total SFR in NGC~7331 ($3.0 \rm \ M_{\odot} \ yr^{-1}$) \citep{Leroy2008}. Alternatively, similar projection effects could also be caused by a warp in the spiral arms. The presence of a complex warp in NGC~7331 has been inferred on large scales from the H\,\textsc{i} distribution and velocity field, and \citealt{Bosma1981} uses the H\,\textsc{i} velocity field to suggest that the northern arm is warped out of the plane. Evidence of a warp is also seen in the inner disk from a gradient in rotation speed between the stars and emission line gas \citep{Bottema1999}. The patch of high-velocity residual may be part of a warped spiral arm ending at the northern edge of the galaxy. 

The central biconical velocity residual seen in Figure \ref{fig:kinematics2} is likely due to beam smearing. The lower resolution H\,\textsc{i} data traces coarser spatial scales, averaging over a larger beam size at the center of the galaxy, where there is a steep velocity gradient. We see a similar effect in the dispersion data, where beam smearing increases the measured velocity dispersion. Additionally, differences in the distribution of neutral and ionized gas at the center of the galaxy could contribute to this residual \citep{Mitchell2015}.

\begin{table*}[ht]
\centering\caption{H\,\textsc{ii} region properties}
\resizebox{\textwidth}{!}{ 
\begin{tabular}{c c c c c c c c c}
\hline\hline 
ID & R.A. & Decl. & Radius 3$\sigma$ & H$\alpha$ Flux & H$\alpha$ Luminosity & SFR (H$\alpha$) & Dispersion & \\
& (J2000) & (J2000) & (arcseconds) & (W/m$^2$) & (W) & (M$_\odot$/yr) & (km/s) & \\
\hline
IPEK001 & 339.2732044 & 34.4149632 & 13.21 & $6.77\times10^{-17}$ & $1.70\times10^{32}$ & 0.0094 & 29.92 \\
IPEK002 & 339.2722128 & 34.4095789 & 21.92 & $6.03\times10^{-17}$ & $1.52\times10^{32}$ & 0.0083 & 28.23 \\
IPEK003 & 339.2573550 & 34.4201773 & 14.00 & $5.41\times10^{-17}$ & $1.36\times10^{32}$ & 0.0075 & 25.54 \\
IPEK004 & 339.2673456 & 34.4268729 & 14.51 & $4.08\times10^{-17}$ & $1.03\times10^{32}$ & 0.0056 & 28.68 \\
IPEK005 & 339.2601370 & 34.4013563 & 13.32 & $3.74\times10^{-17}$ & $9.40\times10^{31}$ & 0.0052 & 23.34 \\
IPEK006 & 339.2629888 & 34.4290239 & 15.96 & $3.74\times10^{-17}$ & $9.40\times10^{31}$ & 0.0052 & 22.88 \\
IPEK007 & 339.2652180 & 34.4280776 & 15.79 & $3.60\times10^{-17}$ & $9.05\times10^{31}$ & 0.0050 & 27.35 \\
IPEK008 & 339.2935695 & 34.3674692 & 19.54 & $3.55\times10^{-17}$ & $8.93\times10^{31}$ & 0.0049 & 25.05 \\
IPEK009 & 339.2749296 & 34.4057000 & 10.63 & $3.36\times10^{-17}$ & $8.44\times10^{31}$ & 0.0046 & 28.91 \\
\ldots & \ldots & \ldots & \ldots & \ldots & \ldots & \ldots & \ldots \\
\hline
\end{tabular}
} 
\tablecomments{Only a portion of this table is shown here to demonstrate its form and content. A machine-readable file containing the full table is available in the online version of the published article.}
\label{table:HII}
\end{table*}

\subsection{Caveats}
\label{sec:caveats}
As mentioned above, beam smearing or the combination of differing line-of-sight velocities due to low spatial resolution can increase the measured velocity dispersion \citep{Epinat2010, Davies2011}. Beam smearing is most severe at high redshifts (low spatial resolutions) and in galaxies with larger inclinations \citep{Davies2011}. Large velocity gradients at the centers of galaxies can also amplify the effects of beam smearing, spuriously increasing the line-of-sight velocity dispersion. To ensure we are not overestimating the H$\alpha$ velocity dispersion, especially in the inner ring of NGC~7331, we calculate the local velocity gradient across each lenslet/spaxel and confirm that there are no spaxels where $v_{g} > 2\sigma$ \citep[e.g.,][]{Bassett2014, Varidel2016, Zhou2017}.

A significant fraction of ionized gas emission occurs beyond H\,\textsc{ii} regions, emanating from the DIG between spiral arms. The DIG has been shown to exhibit large velocity dispersions, and measurements of patchy bright complexes may contain an underlying contribution from this broad diffuse emission \citep{Thurow2005, Oey2007}. Due to our high signal-to-noise cut, we do not expect our spectra to be significantly contaminated by broad, faint emission from the DIG. We leave for future work an examination of broad and multicomponent spectra and lower S/N diffuse regions that may be associated with elevated velocity dispersions. 

NGC~7331 is host to many well-known dust features, with dust distributed throughout the spiral arms and prominent dust lanes seen in the western galactic disk. The central star forming ring is also dusty; it appears bright in $24\mu$m images but is absent in GALEX UV data, likely the result of substantial dust extinction \citep{Thilker2007}. The SINGS NGC~7331 H$\alpha$ map used here is corrected for Galactic foreground extinction but not for intrinsic extinction. For a direct comparison, we do not use the E(B$-$V) extinction corrections for the PHANGS-MUSE and SAMI survey when plotting these datasets in Figure \ref{fig:Lsigma}. Bright H\,\textsc{ii} regions in NGC~7331 that are enshrouded in dust will have a significantly lower observed flux/luminosity. Additionally, an inhomogeneous dust distribution that produces varying intrinsic extinction complicates the interpretation of relative flux/luminosity values for H\,\textsc{ii} regions in spatially distinct environments, such as the disk versus the ring. We acknowledge that uncertainties in the intrinsic reddening correction may weaken the derived correlation between $\rm \sigma$ and $\rm SFR(L_{H\alpha})$, with a stronger correlation often noted between $\rm \sigma$ and $\rm SFR(L_{IR})$ \citep{Arribas2014}. Follow-up observations with CH$\alpha$S in the narrowband H$\beta$ mode \citep{Sitaram2024} would allow us to estimate our own extinction corrections for NGC~7331 and future targets. Corrections for intrinsic extinction have been previously derived in NGC~7331 by \cite{Thilker2007}, finding a global $A\rm_{FUV} = 2.51$ in the disk and $A\rm_{FUV} = 3.3$ in the ring. Following \cite{Thilker2007}, we assume that the dust structure in NGC~7331 is similar to that of the Milky Way. Using the Milky Way extinction curve from \cite{Cardelli1989}, $A\rm_{FUV} = 2.64$$A\rm_{V}$ at $\lambda_{eff} = 1516$ for the GALEX far-ultraviolet (FUV) channel \citep{Morrissey2007} and  $A\rm_{H\alpha} = 0.82$$A\rm_{V}$. Accordingly, H\,\textsc{ii} regions in the disk of NGC~7331 have $A\rm (H\alpha) = 0.78$ and H\,\textsc{ii} regions in the ring have $A\rm (H\alpha)= 1.02$. Accounting for this intrinsic extinction increases the flux/luminosity values by a factor of 2 (0.3 dex) in the disk and up to 2.5 (0.4 dex) in the ring. Accordingly, correcting for intrinsic extinction in Figure \ref{fig:Lsigma} places some of the brightest H\,\textsc{ii} regions on/near the L$-$$\sigma$ envelope, implying they are near virial equilibrium (see Section \ref{subsec:turb}).  
 
\section{Summary} 
\label{sec:summary}

\begin{enumerate}

    \item We investigate the ionized gas kinematics in the H \textsc{ii} regions of the disk of NGC~7331 using IFU spectral imaging collected with the CH$\alpha$S. We present a catalog of 136 H \textsc{ii} regions with radii ranging from  $\sim$150 to 450 pc. Many of these regions fall within the inner ring of dust and gas in NGC~7331, which hosts one-third of the galaxy's current star formation activity.

    \item  High-resolution velocity and dispersion maps of NGC~7331 are presented in this work, selecting spaxels with high signal to noise in order to measure dispersions as low as 12 km s$^{-1}$. Prominent residuals in the center of the $\rm V_{H\alpha} - V_{HI}$ map are likely the result of resolution differences (beam smearing) or variation in the neutral and excited gas distributions. A patch of high residual seen in the northern galactic disk may be an ionized, inflowing cloud seen in projection against the H\,\textsc{i} rotation curve of the galaxy or evidence of a warp in the spiral arm. Follow-up analysis of these features is needed to look for peculiar gas motions.

    \item  The L(H$\alpha$), $\rm \Sigma_{SFR}$ and $\sigma$ measurements we make in NGC~7331 are consistent with spatially resolved observations of H\,\textsc{ii} regions in large surveys of nearby galaxy disks. The dispersion is correlated with the SFR surface density, suggesting that models for intense star formation injecting energy into the ISM are a favorable fit to H\,\textsc{ii} regions in NGC~7331. Using the relation $\rm \sigma \propto \epsilon \Sigma_{SFR}^{\alpha}$ H\,\textsc{ii} regions in NGC~7331 are best fit by  $\epsilon = 80$ , $\alpha =0.285$. The best fit varies slightly but not significantly between the ring and the disk, hindered by scatter. Contributions from the DIG and dust extinction are uncertain and may weaken the derived correlation.
     
    \item En route to ultradeep observations of the circumgalactic medium, CH$\alpha$S will obtain detailed spectral imaging of a large sample of nearby galaxy disks. The methods used in this pilot study will be applied to a larger sample of galaxies observed during the CH$\alpha$S early science campaigns.
    
\end{enumerate}

\begin{acknowledgments}
CH$\alpha$S is funded by NSF AST-1407652. N.M was supported by the Alan Brass Prize Fellowship in Instrumentation and Technology Development. M.S is supported by NASA FINESST 80NSSC22K1603. CH$\alpha$S data were collected at the MDM Observatory, operated by Dartmouth College, Columbia University, Ohio State University, Ohio University, and the University of Michigan. Many thanks to the MDM Observatory staff, Eric Galayda and Tony Negrete, for their help setting up these observations and installing CH$\alpha$S on the Hiltner 2.4 m. Thank you to John Thorstensen and Jules Halpern for their support in scheduling these observations. Thank you to Rob Kennicutt and Sean Linden for the helpful discussions, which have improved this work. 
\end{acknowledgments}

\vspace{5mm}
\facilities{MDM:2.4m, KPNO:2.1m}

\bibliography{sample631}{}
\bibliographystyle{aasjournal}

\end{document}